\documentclass{optica-article}

\journal{opticajournal} 

\articletype{Research Article}

\usepackage{lineno}

\usepackage{graphicx}
\usepackage{dcolumn}
\usepackage{bm}

\newcommand{\ten}[1]{\overline{\overline{#1}}}
\begin{document}

\title{Modeling Atomistically Assembled Diffractive Optics in Solids}

\author{Trevor Kling\authormark{1,*}, Dong-yeop Na\authormark{2}, Mahdi Hosseini\authormark{1,3}}

\address{
\authormark{1}Department of Electrical and Computer Engineering and Applied Physics Program, Northwestern University, Evanston, IL 60208, USA \\
\authormark{2}Department of Electrical Engineering, Pohang University of Science and Technology, Pohang 37673, South Korea \\
\authormark{3}Elmore Family School of Electrical and Computer Engineering, Purdue University, West Lafayette, Indiana 47907, USA
}
\email{\authormark{*}TrevorKling2027@u.northwestern.edu}

\date{\today}

\begin{abstract*}
We develop a model describing long-range atom-atom interactions in a two-dimensional periodic or a-periodic lattice of optical centers considering spectral and spatial broadening effects.  Using both analytical and numerical Green's function techniques, we develop a mathematical framework to describe effective atom-atom interactions and collective behaviors in the presence of disorder. This framework is applicable to a broad range of quantum systems with arbitrary lattice geometries, including cold atoms, solid-state photonics, and superconducting platforms. The model can be used, for example,  to scalably design quantum optical elements, e.g. a quantum lens, harnessing atomistic engineering (e.g. via ion implantation) of collective interactions in materials to enhance quantum properties at scale.  
\end{abstract*}



\section{Introduction}
Recent advances in laser trapping of individual atoms \cite{kaufman2012cooling, barredo2016atom} have sparked interest within the scientific community to investigate the feasibility of constructing single-atom thick optical elements. It has been proposed that a dilute 2D array of optically trapped atoms can be used as a mirror, leveraging the atom-atom interactions to enhance scattering\cite{shahmoon_cooperative_2017, rui_subradiant_2020}. An experimental realization followed shortly thereafter, based on a rectangular array of laser-cooled atoms\cite{srakaew2023subwavelength}.  Further investigation of the coherence of these arrays has revealed the formation of complex optical patterns in simple uniform arrays, even in the absence of a cavity\cite{parmee2023cooperative}.

In principle, these systems are not limited to only reflection, but can also be used to perform diffractive control of light.  The continuously-driven atoms behave like a system of pinholes, allowing for the design of tailored diffraction patterns \cite{kipp_sharper_2001, machado_terahertz_2018}. In the case of laser-trapped atoms, arbitrary control of atomic geometries can be achieved by holographic traps\cite{nogrette2014single} that enables realization of more complex collective interactions. 

Studying arrays of this kind in solids, rather than laser-cooled atoms in vacuum, offers the advantage of scalability, with atom positions controlled by focused ion implantation below the diffraction limit\cite{pacheco2017ion} and with single-ion resolution\cite{pacheco2017ion}. Previously, we developed theoretical and experimental models to study 1D arrays of rare earth ions in solid-state photonic devices\cite{pak2022long, kling2023characteristics, furuya2020study}. In this article, we develop a theory to model the interaction of light with 2D array of atoms (or quantum optical centers) of arbitrary geometries within a solid-state host, accounting for statistical variations in atomic frequency, atom number, and atom locations. With recent advances in ion implantation\cite{pacheco2017ion, titze2021situ}, material interfaces\cite{berkman2023millisecond}, and solid-state spin qubit\cite{zhong2015optically, herbschleb2019ultra} developments, our model allows for design of efficient and large-scale collective effects among millions of optical centers in solids for various applications including quantum storage\cite{lei2023quantum}, entanglement creation\cite{mcconnell2015entanglement} and superradiance interaction \cite{lukin2023two, pak_long-range_2022,white2022enhancing}. 

State-of-the-art implantation techniques allow localization of implanted ions within tens of nanometers of a desired point\cite{titze2021situ, bayn2015generation}, and prior work from our group has shown the emergence of long-range collective behaviors from implanted 1D arrays of rare-earth ions \cite{pak_long-range_2022}.  However, while solid-state systems can support any arbitrary geometry, the implanted ions have their resonant frequencies shifted due to fluctuations in the local crystal field.  The implanted ions form a spectral distribution due to this \textit{inhomogeneous broadening}, which reduces the effective number of atoms at any chosen frequency within the broadened linewidth.  By doping or implanting a large number of these atoms, a similarly dilute array of resonant ions can be produced in a solid-state host.  In this paper, we consider 2D geometries of emitters within a solid-state host and theoretically study collective behavior emerging from long-range atom-atom coupling. We devise a model and develop a mathematical and numerical framework to capture the effects of broadening as, for example, seen in solid-state materials hosting ensemble of quantum centers, e.g. rare-earth ions. The model enables one to study arbitrary atomic geometries with realistic disorder to optimize certain collective properties.

\section{Scattering Dynamics}

\begin{figure*}
    \centering
    \includegraphics[width=1\linewidth]{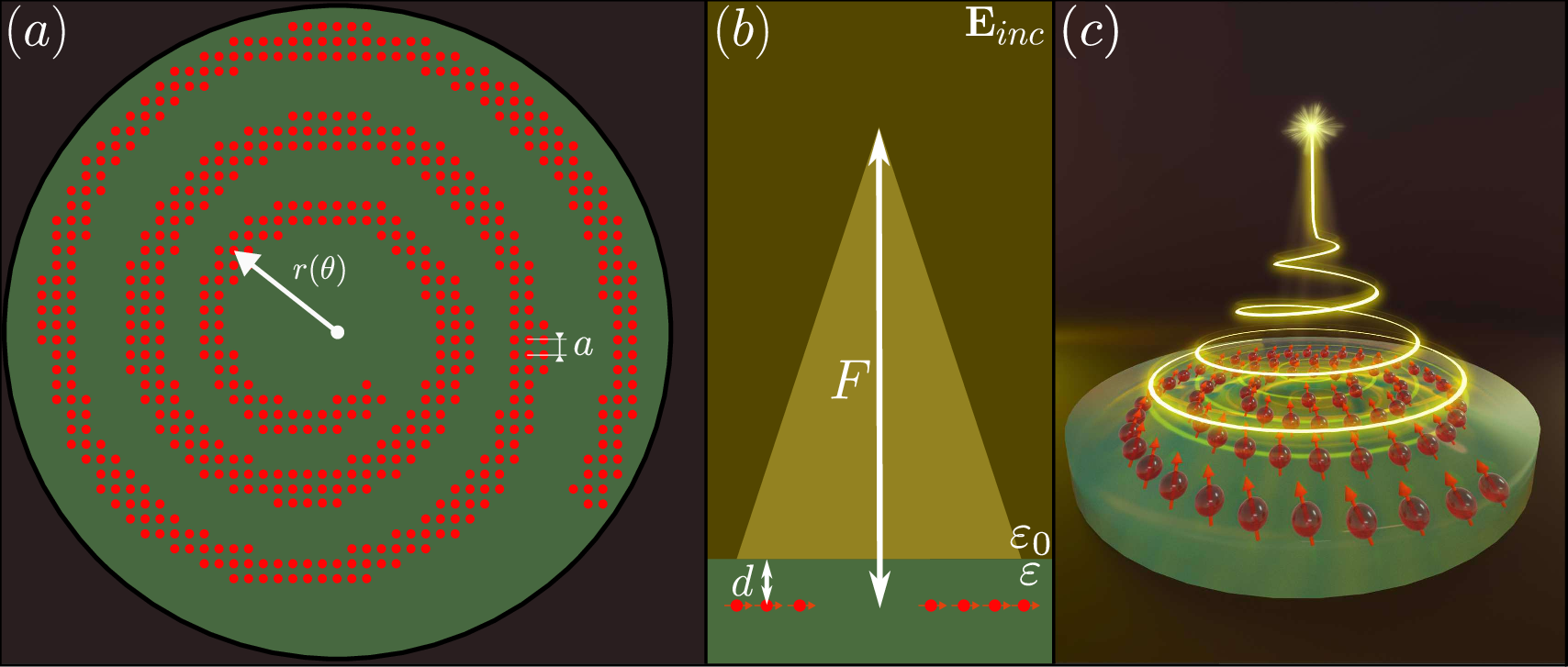}
    \caption{(a) Vertical view: atoms are doped into the substrate in a spiral structure following Eq. \ref{eq:spiral}.  Each implantation region is separated by a distance $a$ in a square lattice, along a spiral determined by Eq. \ref{eq:spiral}.  (b) Horizontal View: the atoms implanted a distance $d$ into the substrate with permittivity $\varepsilon$ are exposed to a uniformly polarized incident field $\mathbf{E}_{inc}$.  The reflected field is then focused a distance $F$ above the atoms, in the free-space region with permittivity $\varepsilon_0$.  (c) 3D view: the phase relationship of the light reflected by a radially-varying set of atoms results in the formation of a hollow beam at the focal point.}
    \label{fig:illustration}
\end{figure*}

Lattice modes have attracted significant prior investigation for their collective properties.  Theoretical works describing these collective behaviors have evaluated the structure of the resulting superradiant and subradiant modes for a number of uniform geometries in one, two, and three dimensions \cite{masson2024dicke, asenjo2017exponential, robicheaux2021theoretical}.  These studies focus on the case of fully ordered atomic systems, as may be achieved by laser trapping atoms. Lattices built based on identical atoms have been shown to be robust to small displacements in their lattice spacing \cite{shahmoon_cooperative_2017}, and as such prior models typically disregard sources of disorder in the system.  However, for a lattice built within a solid-state substrate, intrinsic and fabrication disorder become significant concerns. Understanding the dynamic and static behavior of atomic arrays inside solids is becoming increasingly important given recent advances in scalable solid-state quantum material engineering.  

Atom-atom interactions have been evaluated theoretically and numerically in the case of atomic gasses \cite{ribeiro2024determining, jenkins2016optical}.  Here, the atoms are distributed within a cloud given by a particular probability distribution.  Importantly, these systems demonstrate significant divergence from the mean-field behavior, except in the case of large inhomogeneous broadening \cite{javanainen2017exact}.

In this paper, we develop a model to describe light-atom interactions in an engineered array of solid-state spins or quantum centers. This differs from  prior studies, as the theory captures lattice phenomena as seen in the trapped atom case, but includes the limitations of disorder as seen in the distribution of atoms in a cloud.  This theory establishes itself as a middle ground between the two aforementioned topics; the low-disorder limit the system behaves similarly to a system of trapped atoms, and when the disorder becomes larger than the lattice spacing, the behavior should be similar to that of a randomly distributed gas. Our model can be applied to realistic systems of trapped atoms and solid-state atomic arrays to optimize atomic geometries and enhance specific properties, even in the presence of disorders such as frequency and positional broadening.

For an embedded system, producing arbitrary geometries of quantum centers can be achieved, for example,  by ion implantation techniques.
Here, we construct a spiral atomic system to produce a sort of "atomic lens," where the light scattered by the atoms is focused to a point in the far-field.  This may allow for greater ease of detection of scattering from the atoms, while the single-arm spiral geometry still allows all atoms to couple to common modes propagating along the arm of the spiral.  Other diffractive grating schemes, such as a Fresnel zone plate, may not ensure coherent behavior between spatially separated zones.

For optical applications, we opt to employ a spiral geometry similar to those used in prior studies of hollow beams \cite{jimenez_sharp_2018, fraser2023silicon}, as shown in Fig. \ref{fig:illustration}.  This semi-1D geometry has the benefit of focusing scattered light in the far-field, while maintaining a continuous profile such that all atoms may interact with a common planar mode.  
Tailoring the light to produce such a distribution has applications in atomic trapping.  The geometry of the spiral is determined by the desired focal length $F$ and the wavelength of the light $\lambda$.  We consider the case of a spiral with a single arm, producing a hollow beam with topological charge $+1$.
\begin{equation} \label{eq:spiral}
    r(\theta)^2 = \left(F + \frac{\theta \lambda}{2\pi}\right)^2 - F^2
\end{equation}
Eq. \ref{eq:spiral} provides the parametric equation for a continuous 1D spiral; to translate this behavior to an array of atoms, the spiral is given a finite radial width and atoms are placed with uniform spacing inside the region.  To facilitate focusing, we include a central opaque region by restricting $\theta > 2\pi$.
Locally, we opt to retain the square lattice distribution as in a complete plane.  The atoms are located at positions $\mathbf{r}_n$, determined as the points on a uniform square grid of spacing $a$ that fall within a specified length from the radius of the spiral arm in Eq. \ref{eq:spiral}.  Generally, this could be viewed as ``cutting" the opaque regions of a spiral Fresnel zone plate out of a complete square grid.  We consider exciting the atomic ensemble with linearly polarized light, so maintaining orthogonal symmetry in the local geometry better maintains the polarization of the reflection and increases the cooperative effects.

To observe the optical properties of the system, the atoms are excited by a monochromatic, normally-incident Gaussian beam with uniform polarization in the $x$ direction.

To investigate the behavior of the ensemble under these conditions, we treat our atoms as classical dipoles in the weak-excitation regime, assuming a linear isotropic polarizability \cite{shahmoon_cooperative_2017, svidzinsky2010cooperative, rouabah2014coherence, bettles2015cooperative}.
In this work, we consider a broadened ensemble of atoms such that the polarizability of each atom is dependent on their resonant frequency. 
To describe the behavior of these atoms, we begin with the theory corresponding to an ideal lattice of atoms.  When the system is exposed to an external field, the atoms are driven by both the initial incident field $\mathbf{E}_{inc}$ as well as a scattered field from all nearby atoms $\mathbf{E}_{sc}$.  Depending on the spacing between these atoms, the sum of these fields can be either constructive or destructive, resulting in enhanced or suppressed polarization of the total ensemble.  The polarizations of the atomic dipoles can be found by solving the system of linear equations given in Eq. \ref{polTrancEq}.
\begin{equation}\label{polTrancEq}
    \mathbf{p}_n = \mathbf{p}_{n, inc} + \alpha \frac{k^2}{\varepsilon} \sum_{m \neq n} \ten{G}(k, \mathbf{r}_n, \mathbf{r}_m) \mathbf{p}_m
\end{equation}
Here, $k$ is the angular wavevector of the scattered light, $\varepsilon$ is the material permittivity, and $\mathbf{r}_n$ and $\alpha$ are the position and polarizability of each atom.  $\mathbf{p}_{n, inc}$, the initial polarization, is set by the incident field as $\mathbf{p}_{0, n} = \alpha \mathbf{E}_{inc}(k, \mathbf{r_n})$.  A detailed description of the Green's function is provided in Appendix C, for all cases described in this work.  Each atom is assumed to have a uniform homogeneous linewidth $\gamma$ regardless of its broadened frequency.  To properly include inhomogeneous broadening, the polarizability must allow for variation in both the frequency of the pump and the resonant frequency of the interacting atom.  We define the polarizability of an individual atom in Eq. \ref{eq:alphaDef},
\begin{equation} \label{eq:alphaDef}
    \alpha(\omega_p, \omega_a) = -\frac{3\varepsilon_0 \lambda_a}{n_{ref} k_a^2} \frac{\gamma/2}{(\omega_p - \omega_a) + i\gamma_0/2}
\end{equation}
where $\omega_p$ is the frequency of the pump, $\omega_a$ is the frequency of the interacting atom with corresponding wavelength $\lambda_a$ and wave vector $k_a$, $n_{ref}$ is the refractive index of the material, and $\gamma_{0}$ is the total decay rate \cite{henderson2006optical}.

\begin{figure*}
    \centering
    \begin{minipage}{0.49\linewidth}
        \includegraphics[width=1\linewidth]{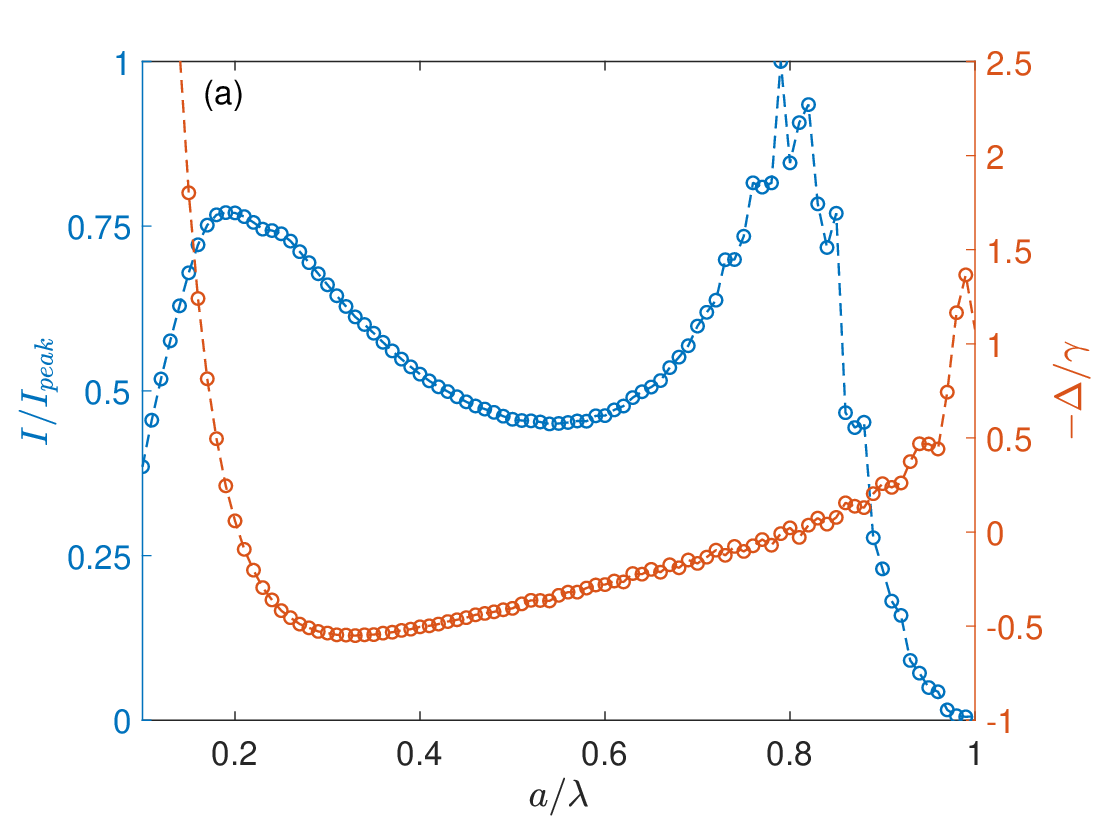}
    \end{minipage}%
    \begin{minipage}{0.49\linewidth}
        \includegraphics[width=1\linewidth]{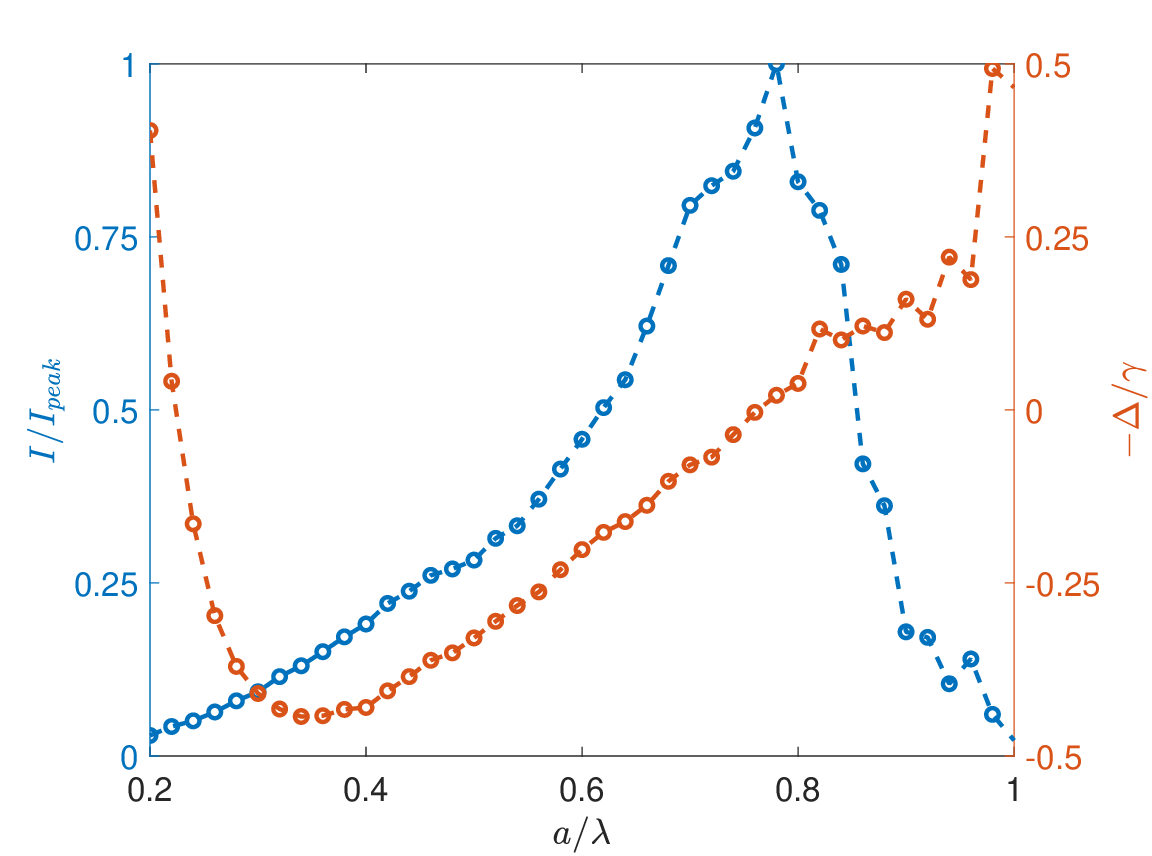}
    \end{minipage}
    \caption{Maximum reflected intensity ($I = \text{max}(|\mathbf{E}_{sc}(\mathbf{r})|^2)$) and ensemble average lattice mode detuning for an ensemble of atoms.  (a) A $50\times 50$ square lattice of atoms with spacing $a$.  The maximum intensity is taken from a region sufficiently far from the plane to eliminate local effects, and is normalized to the highest intensity observed while varying the lattice spacing. (b) A $2000$-atom spiral with a $20 \ \mu$m focal length.  The maximum intensity is taken from the focal region. Numerical simulations were performed with parameters $\{\gamma, \Gamma_{inh}, \omega_a, \varepsilon\} = \{120 \ \text{kHz}, \ 0, \ 377 \ \text{THz}, \ 2.25\varepsilon_0 \}.$}
    \label{fig:freeSpacePosition}
\end{figure*}

\section{Embedded Array}

When the scattering atoms are implanted into a solid-state medium, the behavior of the system is perturbed in three main ways.  Firstly, the introduction of a dielectric surface deflects the scattered light.  This manifests in two effects; a modification to the Green's function corresponding to the reflected light, and a position-dependent phase shift for the transmitted light.  Secondly, atoms implanted into the solid are broadened inhomogeneously due to defects in the host.  This results in a randomization of the phase of the scattering, discussed in greater detail in Section IV.  Finally, implantation procedures have a finite precision in terms of the doping location.  This results in position distribution of the implanted atoms, characterized by a finite uncertainty in the location of the scatterers.

\begin{figure*}
    \centering
    \includegraphics[width=1\linewidth]{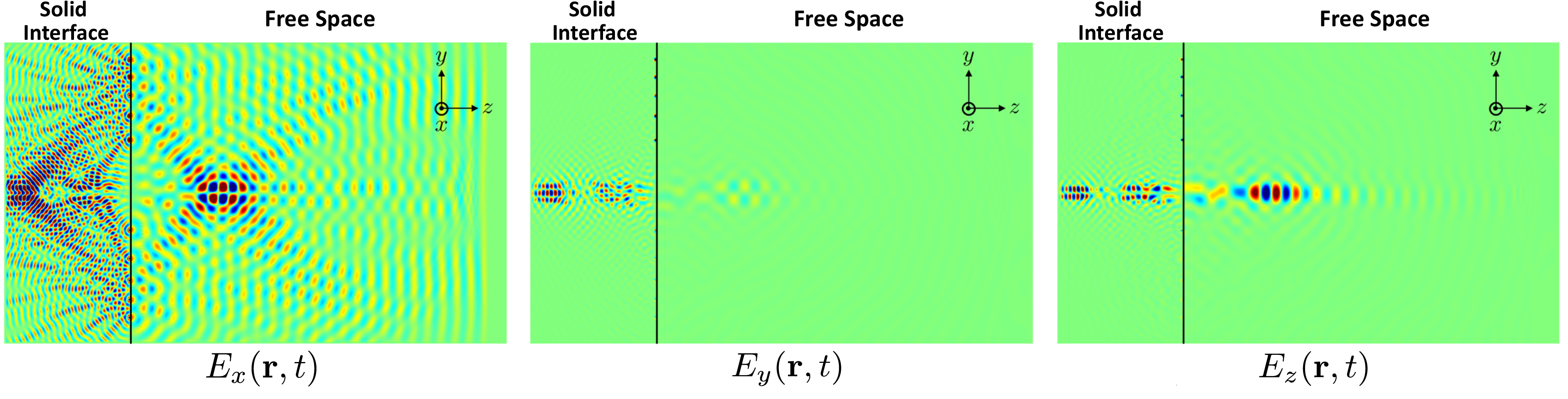}
    \caption{$x,y,z$ components of electric fields obtained from the three-dimensional finite-difference time-domain (FDTD) simulations at time-step index $n=3800$.  The spatial grid size was taken to be $h = \lambda/20$, and each temporal time step spanned a time $\delta t = 0.99 h/c_0$.  Atoms inside the solid host are excited by a $x$-polarized Gaussian beam, and the formation of a hollow beam is observed at a distance $F$ from the array.}
    \label{fig:dielectricSurface}
\end{figure*}

Atoms are implanted in a circular region with radius $\delta r_{max}$, typically on the order of a small fraction of the resonant wavelength.  These implanted atoms undergo inhomogeneous broadening due to local effects from the solid-state host material; as a result, the probability that a given ion has resonant frequency $\omega$ is found via the Gaussian inhomogeneous distribution.  

To account for the inhomogeneous broadening, we consider the effective polarizability of one "site" within the lattice, rather than a singular atom.  Each site contains $N$ atoms with inhomogeneous broadening.  For a pump field at a fixed frequency $\omega$, only a fraction of the implanted atoms will be excited.  To account for this, we define an effective polarizability of each implantation site based on an average of the contribution of each implanted atom.  The distribution of these atoms is a Gaussian, and the average site polarizability can be found in Eq. \ref{eq:avgPol}.
\begin{equation}\label{eq:avgPol}
    \overline{\alpha}(\omega) = \int_{-\gamma}^{\gamma} \alpha(\omega, \omega^\prime) P_{a}(\omega^\prime) d\omega^\prime B(\omega)
\end{equation}
Here, we average over a range of $2\gamma$ to incorporate all atoms with sufficiently strong interaction with the pump field.    The $B(\omega)$ term here is a binomial distribution included to model the effects of radiative quenching.  If multiple atoms with similar frequencies are embedded within the same cite, their large mutual interaction results in radiative quenching.
In typical solid-state systems, e.g. rare-earth ions or defect centers in solids, the homogeneous linewidth is typically much less than the inhomogeneous linewidth.  Using this assumption, the effective polarization of the site is given by \ref{eq:EffPol}, which provides the polarizability of a singular site for any pump frequency $\omega$. 
\begin{equation}\label{eq:EffPol}
    \overline{\alpha}_{di}(\omega) = \alpha(0)\left( \frac{N^2 q(2\gamma)}{4} \left(\frac{2\gamma}{\sqrt{2\pi\sigma^2_{inh}}} e^{-\frac{(\omega-\omega_c)^2}{2\sigma_{inh}^2}}\right)^2 \left(1 - \frac{2\gamma}{\sqrt{2\pi\sigma^2_{inh}}} e^{-\frac{(\omega-\omega_c)^2}{2\sigma_{inh}^2}}\right)^{N-1} \right)
\end{equation}
Here, $\sigma_{inh} = \Gamma_{inh}/2\sqrt{2\ln(2)}$ for the inhomogeneous linewidth $\Gamma_{inh}$, and $\omega_c$ is the central frequency of the inhomogeneous distribution.  $q(2\gamma)$ is a constant corresponding to an integral of the Lorentzian distribution with width $\gamma_0$, integrated over a range $[-\gamma, \gamma]$.  A derivation of this form and further definitions are provided in Appendix B.  This equation can be understood as the resonant response of an effective atom, $\alpha(0)$, multiplied by an effective atom number $\mathcal{N}(\omega)$.

For the following numerical analysis of these systems, the atoms are assumed to be placed in a half-plane of material with a depth $d$. The scattered field is then given by Eq. \ref{eq:scatteredField}, 
\begin{equation}\label{eq:scatteredField}
    \mathbf{E}_{sc}(\mathbf{r}) = \frac{k^2}{\varepsilon}\sum_{m} \ten{G}(k, \mathbf{r}, \mathbf{r}_m)\mathbf{p}_m
\end{equation}
The lattice mode of interest is formed by local interactions between atoms within an arm of the spiral, and direct long-range coupling contributes little to the observed effect.  Instead, collective interaction is facilitated by the lattice mode supported by the sub-wavelength distribution of atoms.  We confirm the existance of a focal region in spite of the dielectric interface via finite-difference time-domain (FDTD) simulations allowing reflections.

For the numerical parameters of the solid, we opt to match our system to a common quantum-optical system of Thulium-doped Lithium Niobate (Tm:LN) surrounded by vacuum. However, the results can be extended to many other ions or emitters in different hosts.  Our choice is motivated by the fact that the production of the kinds of devices simulated here can be achieved using modern nanofabricaiton and implantation techniques on commercially available wafers.

Of great interest in these systems is the effect of interaction between neighboring sites.  A dilute array interacts with both the incident field and the field scattered by all neighboring atoms, constituted by the second term in Eq. \ref{polTrancEq}.  Each of these neighboring sites then contributes an interaction term that constitutes a phase shift and dissipation, dependent on the spatial relationship between the interacting sites.  The coherent sum of all of these interactions acting on a single site produces a cooperative phase shift and linewidth shift, modifying the total effective detuning to $\ten{\delta}_{eff} = \delta_n \ten{I} + \ten{\Delta}_n$ and the total effective linewidth to $\ten{\gamma}_{eff} = \gamma_0\ten{I} + \ten{\Gamma}_n$.  This collective behavior can be given in terms of the Green's function for the scattered field acting on the sites, as shown in Eq. \ref{eq:Interaction}
\begin{equation}\label{eq:Interaction}
    \ten{\Delta}_n + \frac{i}{2}\ten{\Gamma}_n = -\frac{3\gamma \lambda_a}{2 n} \mathcal{N}(\omega)^2 \sum_{m}\ten{G}(k, \mathbf{r}_n, \mathbf{r}_m)
\end{equation}

For comparison, Figure \ref{fig:freeSpacePosition} illustrates the calculated scattering and detuning for free-space square and spiral lattices.  Here, each lattice site hosts exactly one atom at the resonant frequency.  In both cases, the reflected intensity $I$ achieves a maximum around $0.8\lambda$, where the lattice-induced detuning is zero.  The square lattice demonstrates a second local maximum around $0.2\lambda$, which is not present in the spiral case.  For an infinite square lattice, the symmetry of the lattice enhances scattering from the lattice proportional to $\Gamma$ \cite{shahmoon_cooperative_2017}.  This compensates for the reduced polarizability of the ensemble at small lattice spacings, resulting in the second observed peak.

Here, the condition $\Delta/\gamma = 0$ corresponds to the frequency of the incident field being on-resonance with the cooperative resonance produced by the array of atoms.  We note here that this does not imply that the atom-atom interaction vanishes, but rather that the magnitude of the detuning induced by coherent sum of all atom-atom interactions is minimized.  The single-atom parameter $\Delta/\gamma$ is averaged over the complete ensemble, and thus represents a measure of the collective behavior of the lattice.

While two resonances are observed in these systems, the physics of each resonance is distinct.  For small lattice spacing, the high atom-atom interaction leads to superradiant emission, with the cooperative linewidth shift $\Gamma$ diverging as $a$ goes to zero.  Meanwhile, the second resonance at larger atomic spacing has a subradiant characteristic, with $\Gamma < 0$. 

\begin{figure}
    \centering
    \includegraphics[width=1\linewidth]{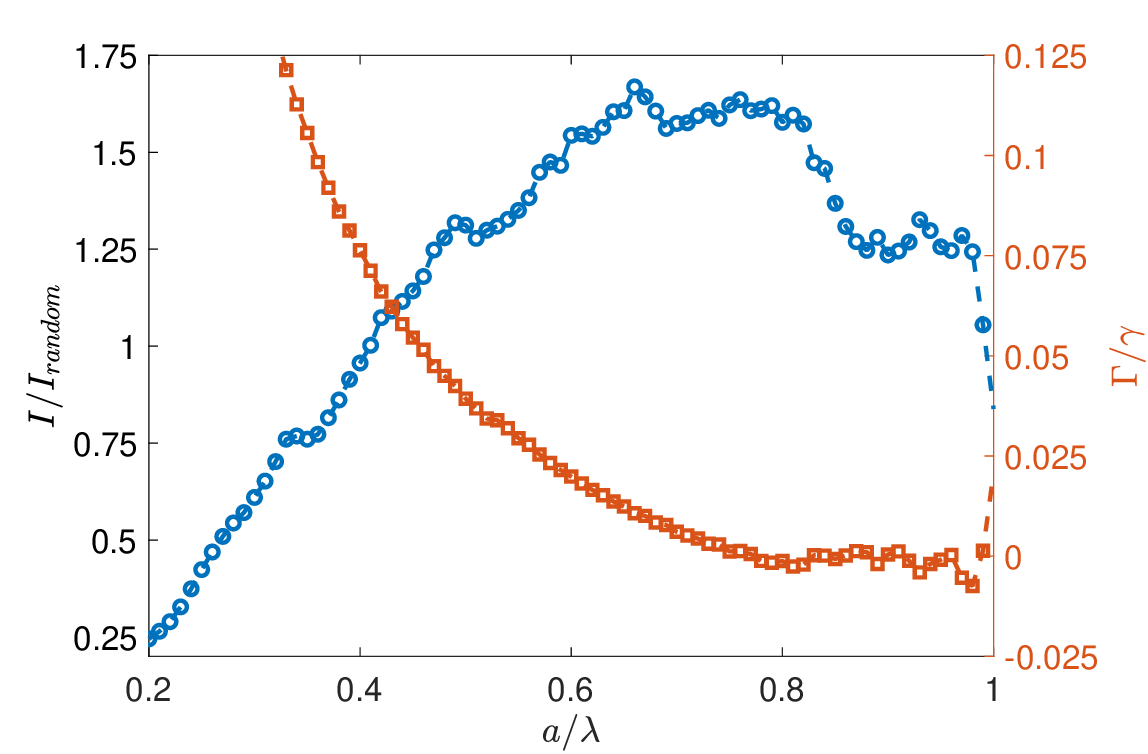}
    \caption{Reflection from an embedded spiral lattice with inhomogeneous broadening and surface reflections versus lattice spacing.  The blue curve indicates the reflected field intensity at the focal point of $20, \mu m$ normalized to the reflection from a spiral with atoms distributed randomly within its arms, $I_{random}$.  The orange curve correspond to the cooperative linewidth shift of the atoms, averaged over the ensemble. Numerical simulations were performed with parameters \(\{F, \gamma, \Gamma_{inh}, \omega_c, \varepsilon\} = \{20 \mu\text{m, } 120 \ \text{kHz}, \ 300 \ \text{GHz}, \ 377 \ \text{THz}, \ 2.25\varepsilon_0\}\).  The pump frequency is taken to be the same as the atomic frequency.  Effects due to nonradiative decay were ignored, taking $\gamma_0 = \gamma$.}
    \label{fig:embeddedScattering}
\end{figure}

In order to observe the focusing effect of electromagnetic waves from a spiral-shaped array of atoms thinly embedded in a dielectric medium, numerical simulations of the scattered electromagnetic field were performed.  Simulations of the scattered field intensity were carried out on the Quest High-Performance Computing cluster at Northwestern University, while a spatially-resolved simulation of the focusing behavior with a dielectric interface were completed on a workstation at Pohang University of Science and Technology (POSTECH), South Korea.   To simulate the effect of an interface on the collective lattice mode, a three-dimensional FDTD algorithm was utilized.  A detailed description of the numerical method is provided in Appendix G.

In order to verify one can retain the diffractive focusing in the presence of a dielectric interface, Figure \ref{fig:dielectricSurface} depicts x, y, and z components of the electric field obtained from the three-dimensional FDTD simulation at time-step index $n=3800$ on the $yz$ plane.  Simulations only involving the focal plane of the structure can be managed efficiently by using the far-field forms of the scattered field.  However, simulations covering all of space must use the near-field equations, which for the approach employed in \ref{eq:scatteredField} involves computing a large number of numerical integrals as well as dealing with the singular nature of the Green's function.  FDTD simulations can be used to find the field distribution and Green's function for an arbitrary distribution of scatters and dielectrics without these constraints, but cannot model the effects of atom-atom interaction.  Here, we demonstrate the focusing behavior of the adjusted site geometry. Since the dipole moment of the atoms is aligned along the x-axis, the x component of the electric field is strongly formed, and it can be observed that a strong focusing spot occurs approximately 20 $\mu$m away from the interface of the dielectric in the vacuum region as expected.

This simulation uses the radial equation derived in Appendix F to avoid defocusing due to the surface.  For a typical Frensel-type diffraction grating, one would expect to observe a family of subfoci at each harmonic of the original focus, $F/2, F/3, F/4, \dots$.  However, the adjustments required to correct the primary focus also have the consequence of suppressing these additional foci as the radial modifications are only constructive for the primary focus.  Due to limitations of the FDTD simulations, these plots disregard the atom-atom coupling considered in the other calculations of focal behavior.  

Varying the spacing of the underlying lattice $a$ allows for control of the lattice mode, as shown in Fig. \ref{fig:embeddedScattering}.  Here, the pump frequency is set to the central frequency of the inhomogeneous distribution.  The maximum reflected intensity is observed in the vicinity of $a \approx 0.8\lambda$.  This corresponds to a maximization of the collective behavior of the array, indicated by the small reduction in the collective decay rate. The collective nature of interaction is perhaps represented by the ensemble-averaged atomic polarization as evaluated in the next section.

\section{Broadening Analysis}

\subsection{Inhomogeneous Linewidth}

\begin{figure}
    \centering
    \includegraphics[width=1\linewidth]{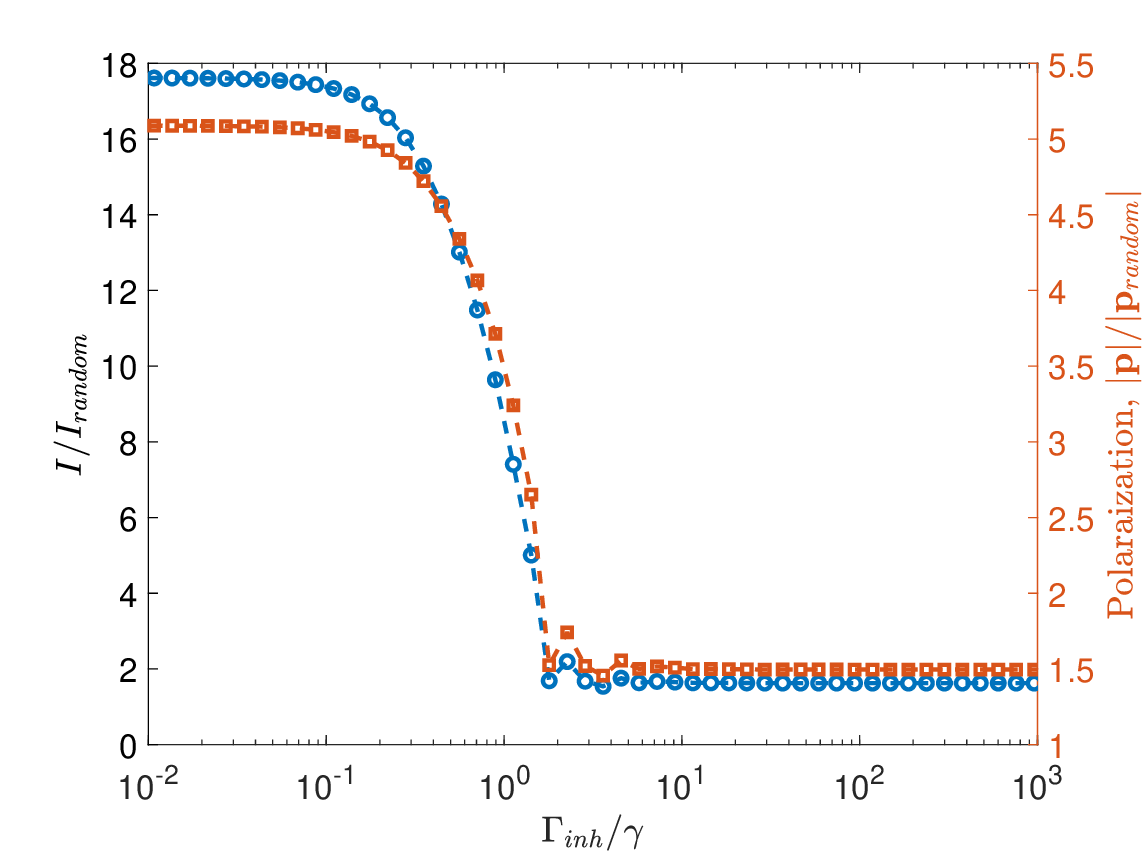}
    \caption{Frequency broadening for a free-space linear spiral with atomic spacing $0.8\lambda$. Numerical simulations were performed with parameters identical to Fig. \ref{fig:embeddedScattering}. Scattering is normalized to a system with equal density of atoms but no lattice structure.}
    \label{fig:simpleBroadening}
\end{figure}

The effect of the inhomogenous linewidth on the collective behavior is examined in Fig. \ref{fig:simpleBroadening}.  Here an ideal array of atoms is given a varying inhomogeneous linewidth, and the reflected intensity and average polarization of the ensemble are simulated.  The number of dopants is varied along with the inhomogeneous linewidth, such that the single-site polarizability is maximized throughout.  The system naturally divides itself into three main regions; one where the inhomogeneous linewidth is much less than the homogenous linewidth, one where the inhomogeneous linewidth is much greater than the homogenous linewidth, and a transitional region where the two are similar in magnitude.

\subsubsection{$\Gamma_{inh} << \gamma_0$}

When the inhomogeneous linewidth is much less than the homogeneous linewidth, the system behaves as an approximately uniform 2D layer of atoms.  The broadening here has little effect on the lattice mode, and the behavior of the system is consistent with prior theoretical analysis such as \cite{shahmoon_cooperative_2017}.  Here, the probability distribution in Eq. \ref{eq:avgPol} approaches a delta distribution as the inhomogeneous linewidth is a fraction of the atomic linewidth.  The behavior of such systems is consistent with ensembles of laser-trapped atoms.

\subsubsection{$\Gamma_{inh} >> \gamma_0$}

When the inhomogeneous linewidth greatly exceeds the atomic linewidth, the system enters into a uniformly-distributed case.  The polarizability of atoms in this scenario was discussed previously in Eq. \ref{eq:EffPol}, where the Gaussian distribution modeling the inhomogeneous broadening is approximately flat over the integration region. This relation is maintained regardless of choice of pump frequency; tuning the pump within the inhomogeneous linewidth can change the effective atom number, but the uniformity of the spectral distribution is generally unchanged.

This case covers the majority of typically-employed doped optical materials.  Defects which affect the degree of inhomogeneous broadening also impact the homogeneous linewidth; as a result, systems with low inhomogeneous broadening typically also have a comparatively narrow homogeneous linewidth.  We note here that the exact width of the inhomogeneous broadening is largely unimportant with regards to the lattice mode, so long as the inhomogeneous width is more than an order of magnitude greater than the homogeneous linewidth.  However, the larger inhomogeneous linewidths require larger dopant concentration in order to achieve equal average population.  As a result, in practice one would expect that system with greater broadening would experience reduced spin coherence and other detrimental effects from crystal damage or defects at high densities.

\subsubsection{$\Gamma_{inh} \sim \gamma_0$}

When the inhomogeneous linewidth is similar to the homogeneous linewidth, the system enters into a distinct behavior regime where the lattice mode has a strong dependence on the pump frequency.  Here, a monochrommatic incident light can couple to a nonuniform distribution of atoms. When the pump frequency matches the central frequency of the inhomogeneous distribution, the system has a symmetric structure with an approximately equal number of positively detuned and negatively detuned atoms excited.  However, by shifting the pump frequency away from resonance, the distribution becomes weighted toward one of the detunings.

\subsection{Spatial Variance}

When the atoms are implanted into a material, the exact location of the atoms has a degree of spatial variation depending on the method of doping and annealing.  This positional variation both influences the strength of coupling between local atoms, as well as shifting the phase of the scattered field.  Increasing this variation results in a incoherent phase background which suppresses the coherent behavior of the lattice, as shown in Fig. \ref{fig:simpleBroadening}.  However, the effects of the spatial variation and inhomogeneous broadening are not necessarily additive; frequency shifts and spatial shifts can result in opposing phase effects, effectively reproducing the ideal lattice behavior.

To interpret the effect of  spatial broadening, we consider a simple system of two implanted atoms separated by a distance $a$ with a uniform circular implantation region of radius $R$.  Consider, for simplicity, that the displacement is along the $x$-axis.  For any pair of displacements in this region $\delta\mathbf{r}_1 - \delta \mathbf{r}_2$, there exist a similar pair with equal radius but opposite change in angle found by mirroring the points across the $x$-axis.  Thus, when the effect of this broadening is averaged over a sufficiently large ensemble, the ensemble-wide change in angle between pairs of points zero.  However, this is not true for the radial displacement.  Consider a semicircular region created by the intersection of a disk centered at the second atom with radius $a$ and the uniform region of implantation centered at the first atom.  Any point on the boundary of this region maintains the radial distance between these points, while any point on the interior reduces it and any point on the exterior extends it.  By inspection, it is clear that there are more points outside the intersection than inside it, and therefore the effect of the uniform distribution is simply an increase to the effective distance between points.  Additionally, this increase in distance is always along the axis of interaction, due to the intersection region being symmetric about its axis.  This applies to both regions; thus, the number of pairs of points where the shifted radius is increased exceeds the number of pairs of points where the radius is reduced.  In Fig. \ref{fig:embeddedScattering}, including spatial broadening shifts the peak observed at a spacing of $0.8\lambda$ back, as the effective distance between points grows.  However, this effect is nonlinear, as the new effective radius depends on both the size of the implantation region as well as the original spacing.  The effect is more pronounced at smaller original spacings, resulting in a broadening of the peak as well.  A diagram of this behavior is provided in Appendix E.

\begin{figure}
    \centering
    \includegraphics[width=1\linewidth]{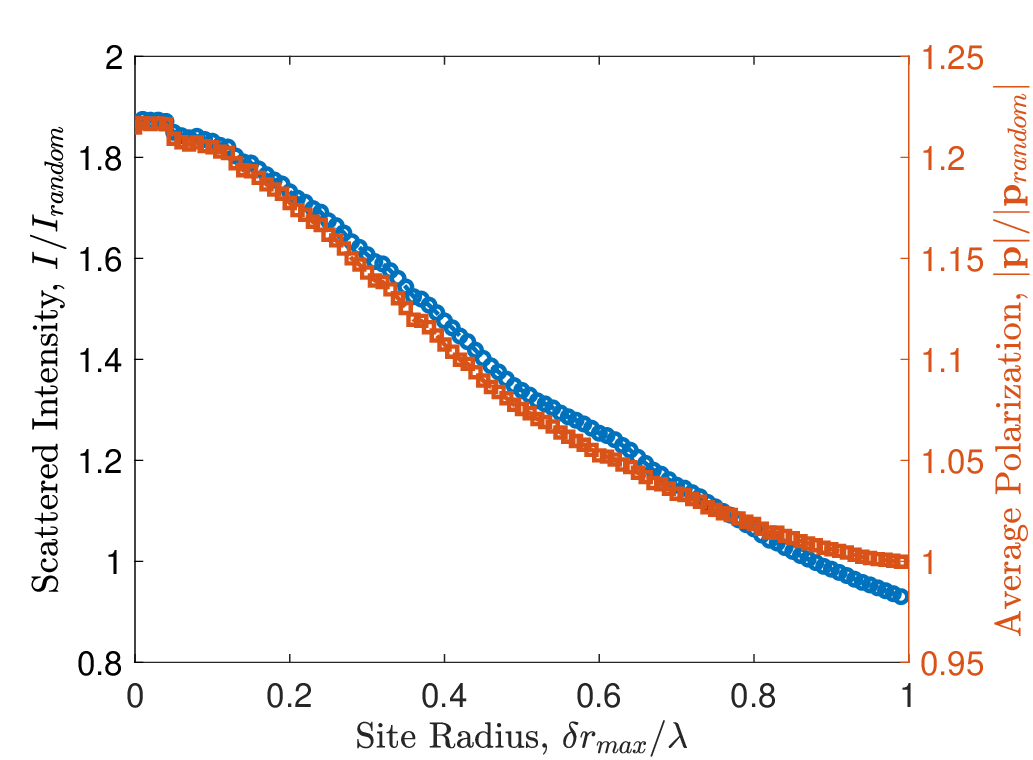}
    \caption[]{Purely positional broadening for an embedded spiral with spacing $0.8\lambda$ and $\Gamma_{inh} >> \gamma_0$.  Numerical simulations were performed with parameters identical to Fig. \ref{fig:embeddedScattering}.  Centers with more than one atom implanted were ignored to avoid changes in the effective number of scatters distorting the trend in the intensity from the lattice mode.  Note that for large site radii, the scattering intensity decreases linearly with radius due to the loss of spatial definition of the diffractive structure (in this case, the spiral).  This effect becomes pronounced around $0.6\lambda$.}
    \label{fig:spatialBroadening}
\end{figure}

\subsection{Dopant Concentration}

The final source of error in these array systems is the number of atoms per region.  When multiple atoms near the same resonant frequency are doped within the same region, they have a much stronger interaction than the engineered atom-atom interaction in the array.  As a result, the phase and intensity of their collective dipole moment is strongly affected; for atoms much less than a wavelength apart, this results in radiative quenching.  For atoms farther apart, the dipoles orient along the common axis.  This produces scattering with highly incoherent phase from these atoms, reducing both the collective effect of the array and the interference pattern in the focal plane.

\section{Discussion}

The above analysis is applicable to a wide range of quantum center arrays. For example, rare-earth ions embedded in a solid-state host such as a lithium niobate slab can be engineered into arrays with arbitrary geometries.  As a multi-functional photonic material, lithium niobate has broad applications in classical and quantum photonics \cite{zhu2021integrated}. Additionally, advances in nanofabrication techniques have resulted in commercially available lithium-niobate-on-insulator (LNOI) wafers, which allow for nanostructures with the aformentioned properties. \cite{dutta2023atomic, lin2020advances}. The integration of active quantum centers such as rare earth ions with LNOI creates additional opportunities for quantum information control and processing\cite{pak2022long, dutta2023atomic, wang2020incorporation, zhao2024cavity}.

It bears mentioning that LNOI is not the only viable host for studying collective effects of rare earth ions; our approach is not specific to  a specific optical center or substrate.  Isotopically pure Silicon is another material with high potential for array fabrication. It possesses much lower inhomogeneous and homogeneous linewidths\cite{gritsch2022narrow, chartrand2018highly}, suitable for observation of relatively strong atom-atom coupling at moderate atomic densities.  We note that for the simulations in this paper, choice of material or quantum centers result in minimal qualitative change to the observed behavior.  However, the reduced inhomogeneous linewidth requires a smaller dosage of ions to resolve the same average ions per site.  As a result, we anticipate that certain experimental properties, such as coherence time, will be improved for a silicon-substrate array compared to LNOI.

In these simulations, we observe a clear lattice resonance from the embedded arrays, even with a large inhomogeneous linewidth.  However, the phase randomization from the change in frequency of the scatterers results in a weaker lattice resonance compared to a free-space case; the lattice mode-induced detuning in the embedded case only achieves 10\% of the effect observed in the zero-broadening designs.  This is caused by the incoherent phase background from the inhomogeneous broadening applied along side the coherent spatial phase.  We note here that we have considered a system with significantly large inhomogeneous broadening; an increased effect may be observed using state-of-the-art low-broadening systems such as isotopically purified silicon or silicon carbide.

At large broadening, one can consider other geometries to improve cooperative behavior. While the collective lattice modes are still observed in the embedded regime, we suggest that a non-uniform spatial distribution may produce even better results.  Prior simulations of free-space systems has shown that the phase shift of a fully detuned ensemble can be compensated for by modifying the lattice spacing \cite{shahmoon_cooperative_2017}.  As such, a sufficiently complex geometry may improve the average behavior by creating an array such that a certain number of nuclei achieve optimal behavior regardless of the single-atom frequencies.  We note here that the focusing effects from a diffractive lens does not require uniform reflection from all points; classical disorganized lenses have been employed in the form of photon sieves for the purposes of broadband performance and control of higher-order modes \cite{zhou_experimental_2009, kipp_sharper_2001}.  Rather, we anticipate a geometry designed with highly nonuniform spacings such that each major region of the spiral would contain some number of atoms with near-ideal couplings.

One avenue to search for such geometries could be the implementation of fractals.  The atom-atom coupling can be interpreted as a form of random walk; such Brownian motion is well-known to produce fractal structures via diffusion-limited aggregation.  A fractal structure may provide the kind of dense phase relationship required to improve performance of these systems in realistic conditions.  Alternatively, one could search the geometric space via machine learning.  Studying properties like the atomic polarization and scattered intensity as optimization targets, one could search for arbitrary 2D geometries that maximize the desired behavior.  Such approaches have been employed to create arbitrary diffraction patterns in traditional optics \cite{huang_ultrahigh-capacity_2015} and to optimize metasurfaces for optics applications \cite{chen2022artificial}.

Such systems could be used as a sort of ``quantum meta-lens". \cite{bekenstein2020quantum}.  With focusing provided by the geometry itself, such systems would have promising applications for squeezed light generation and other nonlinear processes.  Unlike a refractive lens, the focusing here is produced by the phase relationship between the different atoms participating in the scattering.  Thus, under conditions where the ensemble remains coherent throughout its lifetime, one would expect to observe the same focusing effects from the resulting spontaneous emission. To realize such quantum lenses, a focused ion beam \cite{bielejec2010single, pacheco2017ion, titze2021situ} or a nanoscale implantation mask \cite{bayn2015generation} can be used to implant isotopically purified emitters (e.g. rare earth ions) with spatial resolution of 10~nm or less and atom resolution as low as a single atom. The implantation energy can be low to avoid crystal damage, while depth can be controlled by additional material growth after implantation\cite{ding2016multidimensional}. Localizing ions or defects via implantation in hosts with exceptionally low inhomogeneous broadening such as isotopically pure $^{28}$Si \cite{berkman2023millisecond, gritsch2022narrow} or silicon carbide \cite{cilibrizzi2023ultra} opens doors to atomistic design of quantum lenses, and other diffractive optics exhibiting large-scale and scalable collective behavior.

\section{Conclusion}

In this paper, we developed a theory to investigate collective interaction of light with an arbitrary arrays of atoms in a dielectric medium.  Our analytical and numerical method captures realistic broadening effects in solid-state hosts and can be applied to various quantum systems, including a lattice of quantum centers in solid-state photonic materials. We also discuss how materials can be engineered in a top-down approach to harness  collective behavior in atomic arrays and build quantum diffractive optical elements.

\begin{backmatter}

\bmsection{Acknowledgments}
Trevor Kling and Mahdi Hosseini would like to acknowledge support from National Science Foundation Award No. 2410054, and Dong-Yeop Na would like to acknowledge support from the National Research Foundation funded by the Korean Government (MIST), under Grant No. 2023-00213915.
\end{backmatter}

\bibliography{2023ArrayPaperCitations}

\newpage

\appendix
\section{Scattering Formalism Within a Medium}

For simulations of a system of atoms deep within a substrate, the scattering is calculated using the usual Maxwell's curl equation for a monochromatic field in a medium,
\begin{equation}\label{apeq:maxwellCurl}
    \mathbf{\nabla} \times \mathbf{\nabla} \times \mathbf{E}(\mathbf{r}) - \frac{\varepsilon k_0^2}{\varepsilon_0} \mathbf{E}(\mathbf{r}) = \frac{k_0^2}{\varepsilon_0} \mathbf{P}
\end{equation}
where $k_0$ denotes the wavevector in free space.  We note here that we have divided the field-induced polarization into two components; a piece corresponding to the host material which is accounted for by $\varepsilon/\varepsilon_0 = n_{ref}^2$, and a piece from the implanted atoms which is handled by individual dipoles in $\mathbf{P}$.  The solution to this scattering problem can be found in the usual way by considering a intramaterial wavevector $k = n_{ref}k_0$,
\begin{equation}
    \mathbf{E}(\mathbf{r}) = \mathbf{E}_{inc}(\mathbf{r}) + \frac{k^2}{\varepsilon} \int_{V} d\mathbf{r'} \, \ten{G}(k, \mathbf{r}-\mathbf{r'}) \mathbf{P}(\mathbf{r'})
\end{equation}

Ignoring contributions from the interface-induced reflections, the field coupling atoms together and the radiated field from the ensemble can each be represented by the usual free-space Green's function for their repsective wavevectors,
\begin{equation}
    \ten{G}(k, \mathbf{r}-\mathbf{r}') = \Big[ \Big(1 + \frac{i}{kr} - \frac{1}{k^2 r^2}\Big)\ten{\mathbb{I}} + \Big(-1 - \frac{3i}{kr} + \frac{3}{k^2 r^2}\Big)\frac{(\mathbf{r} - \mathbf{r}')(\mathbf{r} - \mathbf{r}')}{r^2} \Big]\frac{e^{i k r}}{4 \pi r}
\end{equation}
where $r = |\mathbf{r} - \mathbf{r}'|$.  When including a surface, this equation is modified by the position dependence of $\varepsilon$.  This effect is explicitly accounted for in the FDTD simulations by solving Eq. \ref{apeq:maxwellCurl} with a piecewise $\varepsilon$, indicating the transition between substrate and free space.

\section{Polarizability}
The polarizability of a single atom implanted in a material with refractive index $n$ is given by
\begin{equation}
    \alpha(\omega_p, \omega_a) = -\frac{6 \pi \varepsilon_0 c^3}{n  \, \omega'^3} \frac{\gamma/2}{(\omega_p - \omega_a) + i\gamma_0/2}
\end{equation}
where $\omega$ is the frequency of the incident light and $\omega'$ is the resonant frequency of the atom.  In the presence of inhomogeneous broadening, an ensemble of atoms take on varying resonant frequencies depending on their local environment.  These fluctuations can be expressed as a Gaussian distribution of the resonant frequencies of the atoms,
\begin{equation}
    P_a(\omega) = \frac{1}{\sqrt{2\pi \sigma_{inh}^2}} e^{-(\omega - \omega_c)^2/2\sigma_{inh}^2}
\end{equation}
where $\omega_c$ is the center of the inhomogeneous distribution.  We consider a system where a region of radius $R$ is implanted with $N$ atoms, uniformly distributed within the region.  We calculate an effective polarizability of this region by considering the probability of observing exactly one atom with a resonant frequency of $\omega'$ within a spectral region of $2\gamma$, constituting an approximate spectral range for the interaction of atoms.
\begin{equation}
    P(\omega, 2\gamma) = \int_{-\gamma}^{\gamma} P_a(\omega -\delta) d\delta
\end{equation}
Having more than one atom within a region of small $R$ results in radiative quenching, inhibiting emissions.  The probability of observing exactly one atom is given by the binomial distribution,
\begin{equation}
    B(\omega) = N P(\omega, 2\gamma) (1- P(\omega, 2\gamma))^{N-1}
 \end{equation}
and the resulting effective polarization is given by
\begin{equation}
    \overline{\alpha}(\omega) = \int_{-\gamma}^{\gamma} N \alpha(\omega, \omega') P_a(\omega') d\omega' B(\omega)
\end{equation}.
Typically, for a dielectric host $\gamma << \sigma_{inh}$.  This allows for $P_a(\omega')$ to be held as constant over the region where $\alpha$ is defined, and the equation to be greatly simplified.
\begin{equation}
    \overline{\alpha}_{di}(\omega) = \alpha(\omega, \omega)\left( \frac{N^2 q(2\gamma)}{4} \left(\frac{2\gamma}{\sqrt{2\pi\sigma^2_{inh}}} e^{-\frac{(\omega-\omega_c)^2}{2\sigma_{inh}^2}}\right)^2 \left(1 - \frac{2\gamma}{\sqrt{2\pi\sigma^2_{inh}}} e^{-\frac{(\omega-\omega_c)^2}{2\sigma_{inh}^2}}\right)^{N-1} \right)
\end{equation}
\begin{equation}
    q(2\gamma) = \int_{-\gamma}^{\gamma} \frac{1}{\delta + i\gamma_0} d\delta
\end{equation}
The contributions here can be divided into two pieces as $\overline{\alpha}(\omega) = \alpha(0)\mathcal{N}(\omega)$; the initial, on-resonant polarizability of a single atom $\alpha(0)$ and a effective atom number which varies with pump frequency $\mathcal{N}(\omega)$.

\section{Green's Functions}
The propagation of fields produced by dipole emitters can be modeled by the electromagnetic Green's function.  In free space, the field emitted by a single dipole emitter with polarization $\mathbf{p}$ at a position $\mathbf{r}'$ and measured a a point $\mathbf{r}$ is given by
\begin{equation}
    \mathbf{E}_{sc}(\mathbf{r}) = \frac{k_0^2}{\varepsilon_0} \overline{\overline{G}}_0(k_0, \mathbf{r} - \mathbf{r}') \mathbf{p}
\end{equation}
where $k_0$ is the wave vector of the emitted light in free space and $\overline{\overline{G}}_0(k_0, \mathbf{r} - \mathbf{r}')$ is the Green's function.
\begin{equation}
    \overline{\overline{G}}_0(k, \mathbf{r}) = \left[ \Big(1 + \frac{i}{kr} - \frac{1}{k^2 r^2}\Big)\ten{\mathbb{I}} + \Big(-1 - \frac{3i}{kr} + \frac{3}{k^2 r^2}\Big)\frac{(\mathbf{r} - \mathbf{r}')(\mathbf{r} - \mathbf{r}')}{r^2} \right]\frac{e^{i k r}}{4 \pi r}
\end{equation}
When considering a dipole embedded near a dielectric interface, this Green's function is modified for each of the half-planes above and below the interface.  We assign values in the lower half-plane an index of 1, while values corresponding to the upper half-plane receive an index of 0.  In the lower half-plane which hosts the emitter, the Green's function has an added term for the reflected waves, given by
\begin{equation}
    \ten{G}_{ref}(k_1, \mathbf{r}, d) = \frac{i}{4\pi}\int_0^\infty dk_\rho \, k_{\rho} \left[ \ten{M}_{ref}^{s} + \ten{M}_{ref}^{p} \right] e^{ik_{z1}(2d - z)}
\end{equation}
where $\ten{M}_{ref}^{j}$ are tensors representing the reflection of $s$- and $p$-polarized waves, respectively.  The system is represented in cylindrical coordinates, with the vector $\mathbf{r}$ rotated to align with the $\theta=0$ axis.  This results in a simple form for the components of the transfer matrices,
\begin{equation}
    \ten{M}_{ref}^{s} = \frac{r^{s}(k_\rho)}{k_{z1} k_{\rho}^2} 
    \begin{bmatrix}
        k_{\rho}^2 (J_0(k_{\rho} \rho) - J_0^{''}(k_\rho \rho)) & 0 & 0 \\
        0 & k_{\rho}^2 J_0^{''}(k_\rho \rho) & 0 \\
        0 & 0 & 0 
    \end{bmatrix} \\
\end{equation}
\begin{equation}
    \ten{M}_{ref}^{p} = \frac{-r^{p}(k_\rho)}{k_1^2 k_\rho^2}
    \begin{bmatrix}
        k_{\rho}^2 J_0^{''}(k_\rho \rho) & 0 & -ik_\rho^3 J_0^{'}(k_\rho \rho) \\
        0 &  k_{\rho}^2 (J_0(k_{\rho} \rho) - J_0^{''}(k_\rho \rho)) & 0 \\
        ik_\rho^3 J_0^{'}(k_\rho \rho) & 0 & -\frac{k_\rho^4}{k_{z1}} J_0(k_\rho \rho) 
    \end{bmatrix}
\end{equation}
The components are given in terms of derivatives of the Bessel function, which can be found through standard recurrence relations. The rotation applied is represented by $\ten{T}(\theta)$, which must be inverted after integration to restore the spatial orientation of the emitted waves relative to the environment.  The transmitted components can be found similarly,
\begin{equation}
    \ten{G}_{tr}(k, \mathbf{r}, d) = \frac{i}{4\pi}\int_0^\infty dk_\rho \, k_{\rho} \left[ \ten{M}_{tr}^{s} + \ten{M}_{tr}^{p} \right] e^{i [k_{z1}(d) + k_{0}(z-d)]}
\end{equation}
\begin{equation}
    \ten{M}_{ref}^{s} = \frac{t^{s}(k_\rho)}{k_{z1} k_{\rho}^2} 
    \begin{bmatrix}
        k_{\rho}^2 (J_0(k_{\rho} \rho) - J_0^{''}(k_\rho \rho)) & 0 & 0 \\
        0 & k_{\rho}^2 J_0^{''}(k_\rho \rho) & 0 \\
        0 & 0 & 0 
    \end{bmatrix} \\
\end{equation}
\begin{equation}
    \ten{M}_{tr}^{p} = \frac{t^{p}(k_\rho)}{k_0 k_1 k_\rho^2}
    \begin{bmatrix}
        k_{\rho}^2 k_{z0} J_0^{''}(k_\rho \rho) & 0 & -ik_\rho^3 \frac{k_{z0}}{k_{z1}} J_0^{'}(k_\rho \rho) \\
        0 &  k_{\rho}^2 k_{z0} (J_0(k_{\rho} \rho) - J_0^{''}(k_\rho \rho)) & 0 \\
        -ik_\rho^3 J_0^{'}(k_\rho \rho) & 0 & -\frac{k_\rho^4}{k_{z1}} J_0(k_\rho \rho) 
    \end{bmatrix}
\end{equation}
Thus, the field emitted by a dipole emitter embedded in a substrate at a depth $d$ is given by
\begin{equation}
    \mathbf{E}_{sc}(\mathbf{r}) = 
    \begin{cases}
        \frac{k_1^2}{\varepsilon} \left[\ten{G}_0(k_1, \mathbf{r} - \mathbf{r}') + \ten{T}(\theta)\ten{G}_{ref}(k_1, \mathbf{r} - \mathbf{r}', d)\ten{T}(\theta)^{-1} \right] \mathbf{p} & z < d \\
        \frac{k_1^2}{\varepsilon} \ten{T}(\theta) \ten{G}_{tr}(k_0, k_1, \mathbf{r} - \mathbf{r}', d) \ten{T}(\theta)^{-1} \mathbf{p} & z > d
    \end{cases}
\end{equation}

\section{Collective Behavior}
To diagnose the collective behavior of our system, we can calculate the eigenmodes of the coupled system.  We consider a simple model of the atom-atom coupling, with a Hamiltonian given by
\begin{equation}
    \mathcal{H}/\hbar = \sum_{n, j} \omega_n \sigma_{ee}^{(n, j)} + \frac{k^2}{\varepsilon} \mathcal{N}(\omega)^2 \sum_{m\neq n, j, l} \mathbf{d}^*_{j} \cdot \ten{G}(k, \mathbf{r}_n, \mathbf{r}_m) \cdot \mathbf{d}_l \, \sigma_{eg}^{(n, j)} \sigma_{ge}^{(m,l)}
\end{equation}
where $n$ and $m$ denote the implantation site index and $j$ and $l$ denote the dipole component. Assuming that our atoms are isotropic, the dipole components can be given by $\mathbf{d}_{j} = |\mathbf{d}|\mathbf{e}_j$, and
\begin{equation}
    \mathcal{H}/\hbar = \sum_{n, j} \omega_n \sigma_{ee}^{(n, j)} - \frac{3\gamma \lambda_a}{2 n} \mathcal{N}(\omega)^2 \sum_{m\neq n, j, l} G_{jl}(k, \mathbf{r}_n, \mathbf{r}_m) \, \sigma_{eg}^{(n, j)}\sigma_{ge}^{(m. l)}
\end{equation}
The collective behavior of the system can be evaluated by finding the eigenvalues and eigenstates of this Hamiltonian; highly collective states will correspond to highly mixed eigenstates, and their eigenvalue will be the corresponding resonant frequency and decay rate modification of the collective mode.  The coefficient of the second term in this Hamiltonian can be reinterpreted as the radiative frequency and linewidth shift of the corresponding atom due to the atom-atom interaction.  Summing over these coefficients gives a measure of the cooperative behavior of the ensemble.

\section{Spatial Variation}

\begin{figure}[!h]
    \centering
    \includegraphics[width=0.5\linewidth]{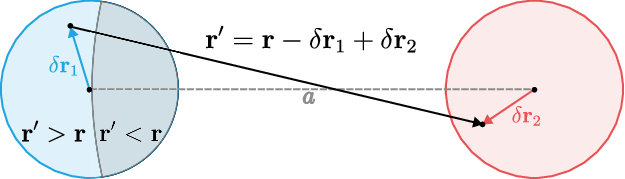}
    \caption{A diagram showing the spatial distributions discussed.  As shown, the region in which the radius is reduced is smaller than the region where the radius is increased.}
    \label{fig:spatialVarianceDiag}
\end{figure}

\section{Surface-Modified Geometry}
When considering dipoles implanted shallowly within a dielectric medium, one additionally must adjust the spiral parameters to account for refraction from the interface.  Specifically, consider a transition between a region of refractive index $n_1$ and $n_2$, with $n_2 > n_1$.  Employing standard ray optics, one can calculate the new path length taken for a source on the dielectric surface at radial distance $R$ to reach a point elevated a distance $F$ above the interface as
\begin{equation}
    L(\theta) = R(\theta)\sqrt{1 + (F/R(\theta))^2} + n_2 d \sqrt{\frac{1 + (F/R(\theta))^2}{\big(1 - (n_1/n_2)^2\big) + \big(F/R(\theta)\big)^2}}
\end{equation}
where $d$ specifies the implantation depth.  The first term here is identical to the path length for the free-space spiral, while the second term gives the adjustment to the path length due to the propagation inside the medium.  This path is then subject to the phase condition,
\begin{equation}
    L(\theta) = F + \frac{\theta \lambda}{2 \pi}
\end{equation}
This equation ensures that, at the focal point $F$, the light from emitters separated by an angular distance of $\pi$ will have a relative phase shift of $\lambda/2$, producing a hollow beam.  Solving this equation numerically for $R$ produces a spiral with varying curvature throughout its angular coordinate, adjusting for the variance in refraction due to changing incident angle.

To then calculate the atom radius $r^{(d)}$ for an implantation depth $d$,
\begin{equation}\label{radialEquation}
    r^{(d)}(\theta) = R(\theta)\Big(1 + \frac{n_1 d}{\sqrt{(n_2 F)^2 + (n_2^2 - n_1^2)R(\theta)^2}}\Big)
\end{equation}
This equation adjusts the atom position to match the surface emitter location to the bulk emitter location, including both angular and optical length considerations.

\section{FDTD Methods}
In order to observe the focusing effect of electromagnetic waves from a spiral-shaped array of atoms thinly embedded in a dielectric medium, 3D electromagnetic numerical simulations were performed. To perform the electromagnetic numerical simulations, a three-dimensional finite-difference time-domain (FDTD) algorithm was utilized.
\begin{figure*}
    \centering
    \includegraphics[width=1\textwidth]{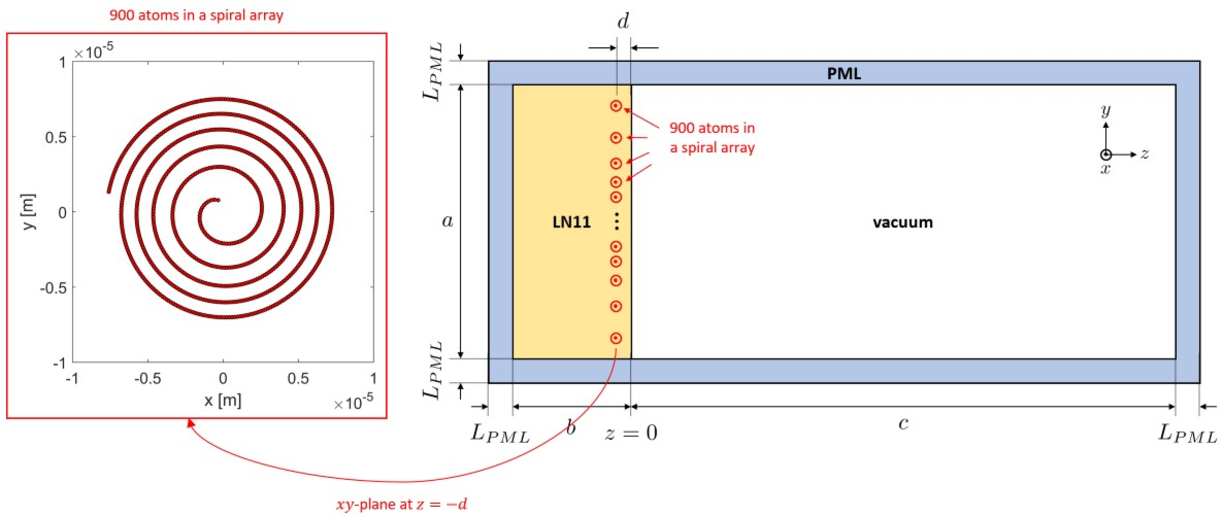}
    \caption{Problem geometry of 900 atoms in a spiral shape on the $xy$-plane embedded in a LN11 dielectric with a depth of $d$ for three-dimensional FDTD simulations. The region of interest including vacuum and LN11 dielectric is surrounded by perfectly matched layers (PML) to model open boundary conditions.}
    \label{fig:simulationdesign}
\end{figure*}
\begin{table}[]
    \centering
    \begin{tabular}{||c|c||c|c||}
        \hline
        $d$ [m] &  $0.1\lambda_{LN11}$ & $\lambda_0$ [nm] & 795 \\
        $a$ [m] &  $28\lambda_{LN11}$ & $\lambda_{LN11}$ [nm] & 352.378 \\
        $b$ [m] &  $20\lambda_{LN11}$ & $h$ [nm] & 17.619 \\
        $c$ [m] &  $60\lambda_{LN11}$ & $\Delta t$ [fs] & 0.0335 \\
        $L_{PML}$ & $10\lambda_{LN11}$ & $n_{LN11}$ & 2.2561 \\
        \hline
    \end{tabular}
    \caption{Parameters used in three-dimensional FDTD simulations.}
    \label{tab:simulationparams}
\end{table}

The problem geometry and specifications are illustrated in Figure \ref{fig:simulationdesign} and Table \ref{tab:simulationparams}. 	The three-dimensional space is divided into a Lithium Niobate 11(LN11) dielectric and a vacuum region. The refractive index of the LN11 material is assumed to be $n_{LN11}=2.2561$, while the refractive index of the vacuum is 1. At a depth, denoted by $d$, from the interface between the LN11 dielectric and vacuum, 900 atoms are embedded. Around the z-axis, these atoms are arranged in a spiral form. Figure 2 shows the arrangement of atoms in the spiral form on the xy-plane. It is assumed that each atom will emit a light with the wavelength of $\lambda_0=795$ nm in free space; hence, the wavelength in the LN11 dielectric becomes $\lambda_{LN11}=352.378$ nm. The average spacing between the atoms is $0.5{\lambda_LN11}$.

\begin{figure}
    \centering
    \includegraphics[width=1\linewidth]{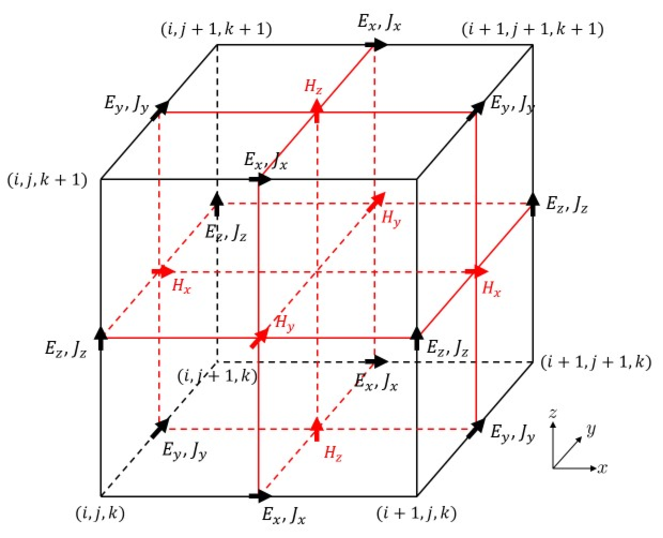}
    \caption{Illustration of the three-dimensional Yee grid.}
    \label{fig:simulationyeegrid}
\end{figure}

\begin{figure}
    \centering
    \includegraphics[width=1\linewidth]{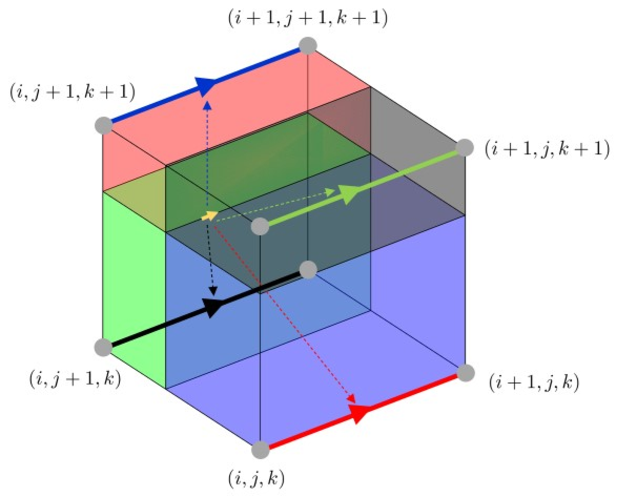}
    \caption{Interpolation of a point current source oriented in the x-direction inside a cubic cell into x-directed edges.}
    \label{fig:simulationunitcell}
\end{figure}

The three-dimensional finite-difference time-domain (3D FDTD) method is a powerful and efficient electromagnetic numerical analysis methodology that can solve Maxwell's equations in time [1]. The space to be analyzed is divided into a number of small cubic cells, and the vector components of the electric and magnetic fields are assigned to the edges and faces of these cells properly known as the Yee grid. The curl operators in Maxwell's equations, which include spatial differential operators, can be approximated using the central difference approximation, allowing for the derivation of equations that represent the interrelationships between the unknown electric and magnetic fields. Additionally, the temporal derivative can also be approximated using the central difference, ultimately enabling the update of late-time unknown vector electromagnetic fields in terms of the addition and subtraction of early-time electromagnetic fields, called the leap-frog scheme (or explicit time-stepping). Figure \ref{fig:simulationyeegrid} illustrates the vector electromagnetic fields and material information assigned to the cubic cells, and the two equations below represent how the 3D FDTD method updates the x-component of the magnetic field in Faraday's law, and the x-component of the electric field in Ampere's law (It applies for the other components by the same token):
\begin{equation}
    H_x |_{i+\frac{1}{2}, j, k+\frac{1}{2}}^{n+\frac{1}{2}} = H_x |_{i+\frac{1}{2},j, k+\frac{1}{2}}^{n-\frac{1}{2}} - \frac{\Delta t}{h \mu_0}\Big(E_x |_{i+\frac{1}{2}, j, k+1}^{n} - E_x |_{i+\frac{1}{2}, j, k}^{n} + E_z |_{i+1, j, k+\frac{1}{2}}^{n} - E_z |_{i, j, k+\frac{1}{2}}^{n} \Big)
\end{equation}
\begin{equation}
    \begin{split}
    E_x |_{i+\frac{1}{2}, j, k}^{n+1} = E_x |_{i+\frac{1}{2}, j, k}^{n} + \frac{\Delta t}{h \varepsilon_r|_{i+\frac{1}{2}, j, k} \varepsilon_0}&\Big(H_y |_{i+\frac{1}{2}, j, k+\frac{1}{2}}^{n+\frac{1}{2}} - H_y |_{i+\frac{1}{2}, j, k-\frac{1}{2}}^{n+\frac{1}{2}} \\& + H_z |_{i+\frac{1}{2}, j, k+\frac{1}{2}}^{n+\frac{1}{2}} - H_z |_{i+\frac{1}{2}, j, k-\frac{1}{2}}^{n+\frac{1}{2}}\Big) - \frac{\Delta t}{\varepsilon_r |_{i+\frac{1}{2}, j, k} \varepsilon_0} J_x |_{i+\frac{1}{2}, j, k}^{n+\frac{1}{2}}
    \end{split}
\end{equation}

where $\Delta t$ and $h$ are the time and spatial grid sizes, respectively. The grid size h is set to be $0.05\lambda_{LN11}$, and $\Delta t$ is set to $0.99h/c_0$ for the stability of the FDTD, where $c_0$ is the speed of light in free space, and 0.99 is the Courant-Friedrichs-Lewy (CFL) condition (should be less than 1). The subscripts $(i,j,k)$ represent grid indices of vertices in the cubic cell, and the superscript n denotes the time index. $E_\xi$, $J_\xi$, and $H_\xi$ represent $\xi(=x,y,z)$ components of the electric field, electric current density, and magnetic field at the respective grid index. In our numerical simulations, as magnetic materials are not considered, the relative permeability of the dielectric is assumed to be 1 over the entire region. The relative permittivity of the dielectric is denoted as $\varepsilon_r$, and $\varepsilon_r |_{i+\frac{1}{2}, j, k}$ represents the relative permittivity assigned at the grid indices $(i,j,k)$. Additionally, to simulate the open boundary conditions, 10 perfectly matched layers (PML) were inserted along each axis on the outer boundary of the three-dimensional analysis region. The size of each PML is equal to $h$.

Each atom was modeled as a tiny dipole polarized in the x-axis direction; hence, it can be modeled as a point current source in the x-direction weighted by the magnitude of the dipole moment in the three-dimensional FDTD simulations. As can be seen in Figure X1, the current density is defined at the edge of the cubic cell. However, the position of atoms in a spiral form does not necessarily lie precisely on the edge of the cubic cell but can be inside the cell; therefore, it is necessary to appropriately interpolate the point current source value to each edge of the cubic cell. Since it was assumed that the dipole moment of the atoms is oriented along the x-axis, the current density interpolation is carried out on the edges in the x-direction of the cubic cell only, and the interpolation for a cubic cell is conducted as follows:

\begin{align}
    J_x |_{i+\frac{1}{2}, j, k}^{n+\frac{1}{2}} &= J_x(x_{atom}, y_{atom}, z_{atom}) \frac{V_{red}}{h^3} \\
    J_x |_{i+\frac{1}{2}, j, k+1}^{n+\frac{1}{2}} &= J_x(x_{atom}, y_{atom}, z_{atom}) \frac{V_{green}}{h^3} \\
    J_x |_{i+\frac{1}{2}, j+1, k+1}^{n+\frac{1}{2}} &= J_x(x_{atom}, y_{atom}, z_{atom}) \frac{V_{blue}}{h^3} \\
    J_x |_{i+\frac{1}{2}, j+1, k}^{n+\frac{1}{2}} &= J_x(x_{atom}, y_{atom}, z_{atom}) \frac{V_{black}}{h^3} \\
\end{align}
where
\begin{align*}
    V_{red} = h(y_{j+1} - y_{atom})(z_{k+1} - z_{atom}), \\
    V_{green} = h(y_{j+1} - y_{atom})(z_{atom} - z_k), \\
    V_{blue} = h(y_{atom} - y_{j})(z_{atom} - z_{k}), \\
    V_{black} = h(y_{atom} - y_{j})(z_{k+1} - z_{atom})
\end{align*}
The current interpolation method is illustrated in Figure \ref{fig:simulationunitcell} more in detail.

\end{document}